\magnification 1200
\centerline {Metastability in Monte Carlo simulation of 2D Ising films and in 
Fe monolayer strips}

\bigskip
Parongama Sen$^{1,2}$, D. Stauffer$^1$, and U. Gradmann$^{3,4}$

$^1$ Institute for Theoretical Physics, 
Cologne University, 50923 K\"oln, Germany

$^2$ Physics Department, Surendranath College, Calcutta 700009, India

$^3$ Physikalisches Institut der TU Clausthal,
Leibnizstr.4, 38678 Clausthal-Zellerfeld, Germany

$^4$ present address Max-Planck-Institut f\"ur Mikrostrukturphysik, 
Weinberg 2, 06120 Halle, Germany 
 
\bigskip
Abstract: 
Effective Curie-temperatures measured in Fe monolayer strips agree reasonably 
with computer simulations of two-dimensional  Ising model strips.
The simulations confirm the domain structure seen already by Albano et al. 

\bigskip
Iron is a ferromagnetic metal, while the Ising model uses only localized spins.
Thus in principle one should not expect good agreement if the Ising model is
compared with experiments on iron. 
Nevertheless, experiments on Fe films and Fe strips [1-3]
could be discussed in terms of 2D Ising model simulations. In particular,
 experiments on Fe(110) monolayer strips prepared on W(110) [1] could
be reasonably well interpreted through Ising model simulations [4]. 
 However, quantitative comparison was not possible both because of incomplete 
information on the Fe monolayer width and because of  limited extent of the 
simulations. Thus we improve the simulations of ref.4 (using a 512 processor 
Cray-T3E instead of a 136 processor Intel Paragon), generalize them to non-zero 
magnetic fields,  and compare them with the old critical temperatures of 
arbitrarily oriented, widely spaced Fe strips  of  incompletely defined width 
[1] as well as with new ones of well oriented, more closely spaced Fe strips  
with better defined  width, which were prepared by step flow growth on miscut 
W(110) substrates [5].

The Monte Carlo simulations, at temperatures below the bulk Curie temperature
$T_c$ of the two-dimensional Ising model (square lattice with nearest-neighbour
interactions between spin 1/2 atoms), were made with standard Glauber kinetics 
and free boundary conditions along the long sides of the $L \times W$ films, 
$L \rightarrow \infty, \; W < 10^2$. In our multi-spin coding method, spins 
are denoted as zero or two instead of the usual zero or one, in order that
the vacuum outside the films can be characterized by one. One Cray-T3E
processor tried to update nearly 14 sites per microsecond. 

In principle, the above $\infty \times W$ geometry of the Ising model does not 
have a sharp phase transition, as is well known: If we wait long enough, the 
magnetization decays
to zero even if (as in our simulations) initially all spins are parallel. In
this sense the strips behave as one-dimensional chains and have a critical
temperature of zero.  However, similar to investigations of glasses,
experiments are made with a finite observation time $1 \dots 10^3 s$  and thus 
may correspond to $10^{12} \dots 10^{15}$ microscopic time units (Monte Carlo 
steps per spin). We thus ask for the temperature $T_c(W)$ at which the 
relaxation time $\tau$ for the magnetization in the Ising strip reaches 
$10^{12}$ or $10^{15}$. This
temperature is only an effective critical temperature; its determination
requires extrapolations via an Arrhenius law, since our simulations are 
restricted to one million time steps.  

Figure 1 shows an example how the magnetization, averaged over many samples,
decays towards zero as exp$(-t/\tau)$ , and how this $\tau$ increases for
decreasing temperature:
$$ \tau = B(W) \exp(A(W)T_c(\infty)/T) \eqno (1)$$
due to an energy barrier $AT_c$ linear in the strip width $W$ which needs to
be overcome. 
Figure 2 gives this roughly linear variation of the energy barrier $A(W)$ with
increasing strip width $W$. We estimate
$$ \tau = 658 \exp[(a T_c(\infty)/T - b)W - 3.31 T_c(\infty)/T] \eqno(2)$$
with $a = 3.13 \pm 0.05, \; b = 3.03 \pm 0.06$. 
Equating this time with $10^{12}$ and $10^{15}$ we find the upper and lower 
curve in figure 3, while the data correspond to ref.1 (larger error bars) and
ref.5 (smaller error bars). We find surprisingly good agreement. 
(The calibration of the W-axis, which could been done in ref. 1 by  a rough 
estimate only, was rescaled by a factor 1.5, compatible with the data of ref. 1,
to better fit both the new experiments and the simulations.
The influence of magnetostatic coupling on Tc in the closely spaced new films 
will be discussed in [5].) However, 
hysteresis experiments [5] for closely spaced films disagree with simulations
in a magnetic field, suggesting that the elementary magnetic dipoles are then
larger than single spins.  

Figure 4
shows the magnetization $M(x)/2$ for $0 < x < L = 10^4$ defined as the number 
of up spins across the narrow strip of width 20. We add suitable constants to 
$M$ to make the whole length $L = 10,000$ visible in ten strips of length 1000
each; these constants give the horizontal lines in figure 4 corresponding to 
16 up spins and 16 down spins.  We see that as in [6], large domains 
separated by relatively sharp domain walls are formed, similar to the time
dependence in the ``tunneling'' of the magnetization [7].  However, the size
of the domains varies appreciably, mostly between $10^2$ and $10^3$. 

These domains now could be regarded as our basic magnetic units but it would be
wrong to identify them with our Ising spins. Instead they form a one-dimensional
chain, with dipole-dipole interactions within one chain and, for closely 
spaced iron films [5], also dipolar interactions from one chain to the 
neighbouring chains. Because of the long range of dipolar forces, such a system
might be described well by a mean field approximation, and such a theory [5] 
will be presented together with the new experiments. It may give even a sharp 
phase transition for infinitely long waiting times, due to the coupling between
different chains.  

In summary, we found a surprisingly good qualitative agreement between widely
spaced iron strips [1] and two-dimensional Ising models, while for closely 
spaced iron films [5] the Ising model without magnetic dipolar interactions
is too simple. 

We thank H.J. Elmers for pointing out an error in a draft of the manuscript,
PS and DS thank SFB 341 for support,  UG the Max-Planck-Institut f\"ur 
Mikrostrukturphysik Halle for hospitality.
\bigskip
\vfill \eject
\parindent 0pt
[1] H.J. Elmers, J. Hauschild, H. H\"oche, U. Gradmann, H.Bethge, D. Heuer,
U. K\"ohler, Phys.Rev.Lett. 73 (1994) 898.

[2] C.H. Back, C.H. W\"ursch, A. Vaterlaus, U. Ramsperger, U. Maier , D. Pescia
Nature 378 (1995) 597

[3] J. Shen, R. Skomski, M. Klaua, H. Jenniches, S. Sundar Manoharan, J. 
Kirschner, Phys. Rev. B, in press (Aug. 97)

[4] D. Wingert, D. Stauffer, Physica A 219 (1995) 141

[5] H.J. Elmers, J. Hauschild, U. Gradmann, in preparation

[6] E.V. Albano, K. Binder, D.W. Heermann, W. Paul, Z.Physik B 77 (1989) 445; 
see also K. Binder, P. Nielaba, V.Pereyra, preprint (1997)

[7] H. Meyer-Ortmanns, T. Trappenberg, J.Stat.Phys. 58 (1990) 185 

\bigskip
Figure captions:

Fig.1: Part a shows the decay of the magnetization with time, part b the 
resulting relaxation time as a function of temperature (Arrhenius law). 

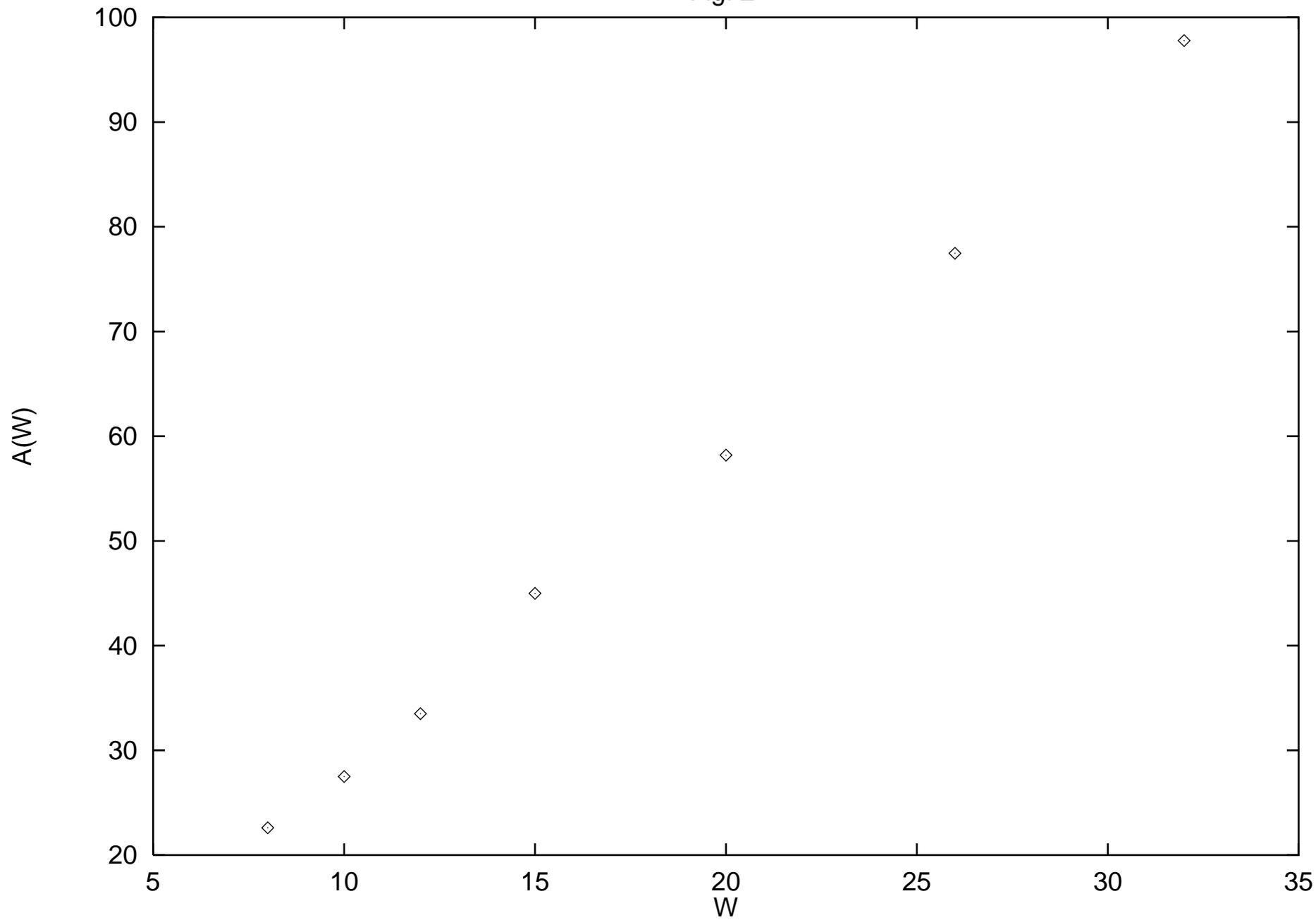
Fig.2: Dependence of Arrhenius parameter $A$, eq.(1), on strip width $W$.

Fig.3: Effective transition temperature (curves) where $\tau$ from eq.(2)
reaches $10^{12}$ and $10^{15}$. 
The bars indicate the width of the transition, not an error of 
measurement.  Wide transitions from widely 
spaced strips [1] have large ``error'' bars, and narrow transitions from 
closely spaced strips [5] have small ones.

Fig.4: Snapshot of equilibrium domain distribution at $T/T_c(\infty) = 0.90,
10000 \times 20, t = 10^6$.  
\end